\documentclass[
  aps,
  pra,
  reprint,
  superscriptaddress,
  floatfix,
  amsmath,
  amssymb
]{revtex4-2}

\usepackage{graphicx}

\IfFileExists{arxiv/supplement_body.tex}
  {\newcommand{\arxivfigdir}{arxiv/figures}}
  {\newcommand{\arxivfigdir}{figures}}
\usepackage{siunitx}
\usepackage[hypertexnames=false]{hyperref}
\usepackage{xcolor}
\usepackage{booktabs}
\newcommand{\suppTableStyle}{%
  \fontsize{9}{10.2}\selectfont
  \setlength{\tabcolsep}{4pt}%
  \renewcommand{\arraystretch}{1.15}%
}
\usepackage{placeins}
\hypersetup{hidelinks}

\usepackage{etoolbox}

\makeatletter
\newcommand{\ba@authornoteseparator}{\textsuperscript{,\,}}

\patchcmd{\@author@join@}
  {\@join{\@separator}{#2}{#3}}
  {\@join{\ba@authornoteseparator}{#2}
    {\frontmatter@footnote{#3}}}
  {}
  {\PackageError{ba-author-notes}
    {Cannot patch author-note storage}{}}

\patchcmd{\doauthor}
  {\frontmatter@footnote{#2}}
  {#2}
  {}
  {\PackageError{ba-author-notes}
    {Cannot patch author-note output}{}}

\pretocmd{\titleblock@produce}{%
  \global\let\ba@saved@title\@title
  \global\let\ba@saved@AAC\@AAC@list
  \global\let\ba@saved@AFF\@AFF@list
  \global\let\ba@saved@AFG\@AFG@list
  \xdef\ba@saved@affilcount{\the\c@affil}%
}{}{\PackageError{ba-supplement-title}{Cannot save REVTeX author data}{}}
\pretocmd{\frontmatter@footnote@produce}{%
  \global\let\ba@saved@FMN\@FMN@list
}{}{\PackageError{ba-supplement-title}{Cannot save author-note marks}{}}

\let\ba@supp@footnotemark\frontmatter@footnotemark
\patchcmd{\ba@supp@footnotemark}{frontmatter.}{supp.frontmatter.}
  {}{\PackageError{ba-supplement-title}{Cannot patch SI note links}{}}
\let\ba@supp@makefntext\frontmatter@makefntext
\patchcmd{\ba@supp@makefntext}{frontmatter.}{supp.frontmatter.}
  {}{\PackageError{ba-supplement-title}{Cannot patch SI note anchors}{}}
\newcommand{\makesupplementtitle}{%
  \title@column{\ba@supplementtitleblock}%
}
\newcommand{\ba@supplementtitleblock}{%
  \begingroup
    \normalsize\frontmatter@setup
    \def\@mpfn{mpfootnote}%
    \def\thempfn{\thempfootnote}%
    \c@mpfootnote\z@
    \let\@makefnmark\frontmatter@makefnmark
    \let\@AAC@list\ba@saved@AAC
    \let\@AFF@list\ba@saved@AFF
    \let\@AFG@list\ba@saved@AFG
    \let\@FMN@list\ba@saved@FMN
    \c@affil\ba@saved@affilcount\relax
    \let\frontmatter@footnotemark\ba@supp@footnotemark
    \let\frontmatter@makefntext\ba@supp@makefntext
    \let\frontmatter@footnote@produce\frontmatter@footnote@produce@footnote
    \phantomsection
    \label{sec:arxiv-supplement}
    \begingroup
      \frontmatter@title@above
      \frontmatter@title@format
      Supplemental Material for ``\ba@saved@title''\par
      \frontmatter@title@below
    \endgroup
    \groupauthors@sw
      {\frontmatter@author@produce@group}%
      {\frontmatter@author@produce@script}%
    \frontmatter@RRAPformat{%
      \produce@RRAP{Dated: \today}
    }%
    \par\frontmatter@finalspace
  \endgroup
}
\makeatother

\begin{document}

\title{Rapid high-purity loading of barium ion via relayed selectivity from naturally abundant source}

\author{Zichun Zhou}
\thanks{These authors contributed equally to this work.}
\email{zczhou24@m.fudan.edu.cn}
\affiliation{Department of Physics, Fudan University,
Shanghai 200433, China}

\author{Xuhui Yao}
\thanks{These authors contributed equally to this work.}
\affiliation{Department of Physics, Fudan University,
Shanghai 200433, China}

\author{Jiaze Li}
\affiliation{Department of Physics, Fudan University, Shanghai 200433, China}

\author{Yuncheng Xie}
\email{xyc271828@gmail.com}
\affiliation{Department of Physics, Fudan University, Shanghai 200433, China}
\affiliation{Key Laboratory for Computational Physical Sciences (MOE) and Institute of Computational Physical Sciences, Fudan University, Shanghai 200433, China}

\author{Haozhe Shi}
\affiliation{Department of Physics, Fudan University, Shanghai 200433, China}
\affiliation{Key Laboratory for Computational Physical Sciences (MOE) and Institute of Computational Physical Sciences, Fudan University, Shanghai 200433, China}

\author{Yanxin Liu}
\affiliation{Department of Physics, Fudan University, Shanghai 200433, China}

\author{Weibin Chu}
\affiliation{Department of Physics, Fudan University, Shanghai 200433, China}
\affiliation{Key Laboratory for Computational Physical Sciences (MOE) and Institute of Computational Physical Sciences, Fudan University, Shanghai 200433, China}
\affiliation{State Key Laboratory of Surface Physics, Fudan University, Shanghai 200438, China}

\author{Lei Feng}
\email{leifeng@fudan.edu.cn}
\affiliation{Department of Physics, Fudan University, Shanghai 200433, China}
\affiliation{Institute of Micronano Electronic Devices and Quantum Computing and Shanghai Key Laboratory of Meta-surface Optoelectronic Fine Control, Fudan University, Shanghai 200438, China}
\affiliation{Hefei National Laboratory, Hefei 230088, China}

\date{\today}

\begin{abstract}
Trapped ions is one of the promising platforms for realizing quantum information processing and quantum simulation. Selective and rapid loading of a specific isotope lays the critical foundation for such purpose particularly when scaling up the system size. Here we demonstrate a new method for rapid isotope-selective loading of
$^{137}\mathrm{Ba}^{+}$ or $^{135}\mathrm{Ba}^{+}$, two types of promising ion qubits, from naturally abundant source using a
three-step, ultraviolet-free photoionization scheme. Neutral atoms are
selected on the narrow \SI{791}{nm}
$6s^2\,{}^1S_0\rightarrow6s6p\,{}^3P_1^o$ transition, resonantly transferred at \SI{554}{nm} to $6s6d\,{}^3D_1$, and ionized using an
existing \SI{532}{nm} Raman beam.
A spectral-overlap analysis based on isotope and hyperfine shifts shows relayed selectivity that the second step can further suppress undesired-isotope excitation and no contaminant ion is detected during our entire experiment.
Besides, the resonant drive of the second step excitation yields a prominent loading rate of the selected species.
This strategy using multi-step excitations to establish relayed selectivity can be extended to high-purity loading of other species of ions.
\end{abstract}

\maketitle

Trapped ions is an attractive platform for quantum
information processing and quantum simulation \cite{RevModPhys.75.281,RevModPhys.93.025001}. Various atomic ion species have been demonstrated as good candidates
for qubits \cite{Debnath2016,Schafer2018,Ransford2026}.
Barium isotopes, $^{137}\mathrm{Ba}^{+}$, $^{135}\mathrm{Ba}^{+}$ and $^{133}\mathrm{Ba}^{+}$, show great promise because they have rich level structures and optical transitions in the visible and infrared ranges. Nonzero nuclear spin provides suitable hyperfine qubit encoding and the
long-lived $5D_{5/2}$ manifold provides optical and metastable-state qubits and motivates the optical--metastable--ground (\textit{omg}) architecture
\cite{Allcock2021}. This also enables high-dimensional qudit control
\cite{Low2025,Low2026}.
Particularly the goldilocks atomic ion of $^{133}\mathrm{Ba}^{+}$ with nuclear spin $I = 1/2$ is ideal for hyperfine qubit \cite{Hucul2017,Christensen2020}.
However, it is radioactive and only available from nuclear synthesis reaction. In contrast, $^{137}\mathrm{Ba}^{+}$ and  $^{135}\mathrm{Ba}^{+}$ with nuclear spin $I = 3/2$ are stable isotopes but come with a more complicated scheme for high-fidelity state preparation and detection \cite{Dietrich2010,An2022}.

Rapid and selective loading of barium ions is a practical challenge for scaling up the number of qubits, particularly from natural-abundance sources. Two-step ionization is a leap forward from the early time electron bombarding or single ultraviolet (UV) photon ionization.
The simplest two-photon approach is just to use a single \SI{413}{nm} laser to first drive the $6s^2\,{}^1S_0\rightarrow5d6p\,{}^3D_1^o$ transition and ionize using the same laser \cite{Leschhorn2012}. It is cost-effective, but its isotope selectivity is limited by spectral overlap and efficiency limited by relatively low excitation and ionization strengths \cite{Leschhorn2012,Choi2024}. An alternative scheme is to drive the strong transition of $6s^2\,{}^1S_0\rightarrow6s6p\,{}^1P_1^o$ with a \SI{554}{nm} laser and use another \SI{405}{nm} laser for ionization \cite{White2022}. Further improvement of efficiency can be achieved by driving an autoionizing transition for the second step \cite{Greenberg2024,Wall2026}. This enhances ionization efficiency but not the isotope discrimination. Loading from a magneto-optical trap (MOT) is fast and isotope-selective but requires extra costly, complex apparatus for neutral-atom cooling and transport \cite{Sage2012,Johansen2022}.

In comparison, the less strong $6s^2\,{}^1S_0\rightarrow6s6p\,{}^3P_1^o$ transition near
\SI{791}{nm} is particularly favorable for isotope selection because of its
well-resolved isotope-shifted hyperfine structure. Selective loading using this transition has been demonstrated with a second UV ionization laser \cite{Steele2007,Wang2011}. However, UV laser is not cost-effective and scattered UV photons induce charging on the ion trap \cite{Harlander2010}.
To overcome this limitation and further improve the loading rate, UV-free three-photon schemes are proposed and studied using \SI{791}{nm} excitation followed by a second resonant transition before the final ionization \cite{Graham2014,Lewty2014, SM}.
Similar multistep photoionization strategies have also been explored in other alkaline-earth species, including Ca and Sr \cite{CPX2017,Cheon2018}.

\begin{figure}[!t]
\vspace{0pt}
\centering
\includegraphics[width=\columnwidth]{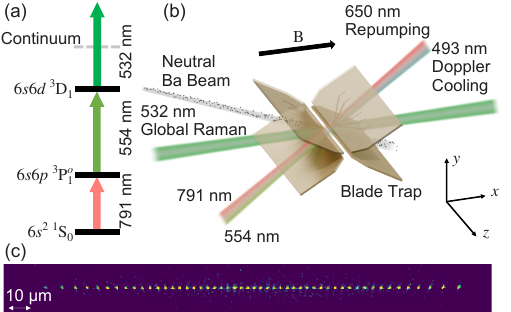}

\caption{Three-step photoionization.
(a) In this work, naturally abundant neutral Ba is first selectively excited through the \SI{791}{nm} transition, then resonantly
transferred by the \SI{554}{nm} laser to $6s6d\,{}^3D_1$, and finally ionized using the \SI{532}{nm} global Raman beam.
(b) Geometry of the neutral atomic beam and the laser beams at the ion
trap. The near co-propagating \SI{791}{nm} and \SI{554}{nm}
beams and the \SI{532}{nm} global Raman beam intersect at the trap
center. The \SI{493}{nm} Doppler-cooling and \SI{650}{nm} repumping
beams, together with the applied magnetic-field direction,
are shown for reference.
The oven, its exit slit, and the \SI{614}{nm} deshelving beam
applied during loading are omitted for clarity.
(c) A 51-ion $^{137}\mathrm{Ba}^{+}$ Coulomb crystal loaded in
\SI{10}{s} from natural-abundance Ba, in which $^{137}$Ba constitutes
only 11.23\%.}
\label{fig:scheme}
\end{figure}

Here we experimentally demonstrate a three-step photoionization scheme comprising excitation $6s^2\,{}^1S_0\rightarrow6s6p\,{}^3P_1^o$ at \SI{791}{nm}, further resonant transfer at \SI{554}{nm} to the $6s6d\,{}^3D_1$ state, and photoionization at \SI{532}{nm}.
We point out that this scheme, via relayed selectivity from the first two steps, can achieve extremely high-purity loading.
A similar scheme was first mentioned in Ref.~\cite{Dietrich2009}, but neither the second-step transition nor the resulting isotope selectivity was discussed in detail.
In our work, we use first-principles calculations to predict
the second-step level structure and guide the search for
$^{137}$Ba and $^{135}$Ba loading resonances
\cite{SM}. We next comprehensively characterize the \SI{791}{nm} loading spectrum combined with the enhancement associated with the second resonant
\SI{554}{nm} transition, and the \SI{532}{nm} ionization-power
dependence.
The last step that drives electron from $6s6d\,{}^3D_1$ to continuum can be achieved with any laser with a energy above the threshold that lies close to \SI{882}{nm}.
\SI{532}{nm} is chosen because it is already employed for stimulated-Raman control in our apparatus, thus can be easily replaced by other convenient choices.
This hardware-efficient implementation also avoids ultraviolet
illumination, thereby eliminating a common source of photoinduced
charging in ion traps \cite{Wall2026,Harlander2010}.

\begin{figure}[!b]
\vspace{0pt}
{%
  \includegraphics[width=\columnwidth,
    trim=0 0 0 0,
    clip]{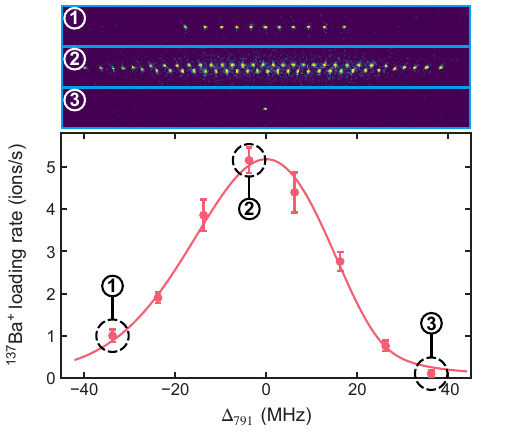}%
}
\vspace{-2.5em}
\caption{Measured $^{137}\mathrm{Ba}^{+}$ loading rate versus
\SI{791}{nm} laser detuning, $\Delta_{791}$, referenced to the peak of
the fitted curve. Each point is the mean loading rate from five
\SI{10}{s} loading trials. Error bars denote one standard error
of the mean. The solid curve is a fit to the conditional slit- and
velocity-averaged excitation model, with a FWHM of
$36(1)~\mathrm{MHz}$ and a signed departure from perpendicular
incidence of $\theta=-5.1(2)^\circ$. This angle is conditional on
the model, rather than an independent alignment measurement
(see Supplemental Material \cite{SM}). All 997 ions loaded during the scan are identified
as $^{137}\mathrm{Ba}^{+}$. The top inset shows exemplary experiment shots, recorded at the correspondingly labeled detunings in the spectrum.}
\label{fig:loading791}
\end{figure}

\begin{figure*}[!t]
\centering
\includegraphics[width=\textwidth]{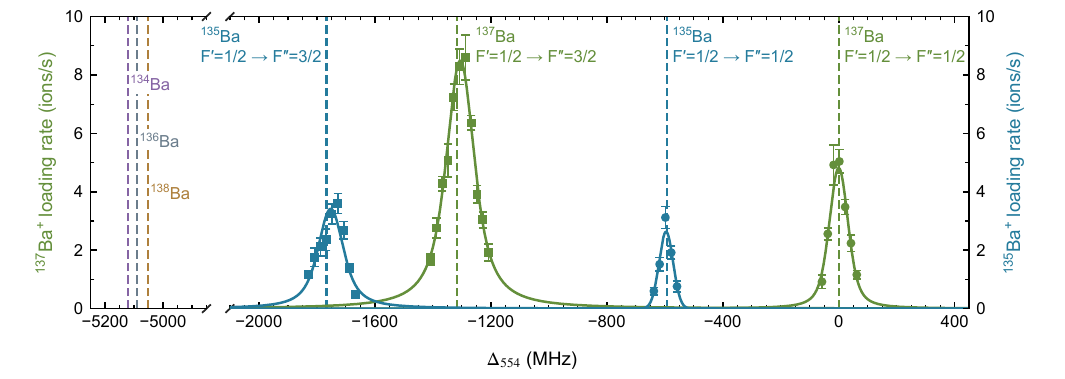}
\vspace{-2.0em}
\caption{
\SI{554}{nm} loading spectra of $^{137}\mathrm{Ba}$
(green, left axis) and $^{135}\mathrm{Ba}$ (blue, right axis),
with identical vertical scales.
For each isotope, the \SI{791}{nm} laser addresses its
$F'=1/2$ component while the \SI{532}{nm} beam is held fixed.
Circles and squares denote the assigned
$F'=1/2\rightarrow F''=1/2$ and $F'=1/2\rightarrow F''=3/2$ branches, respectively.
Each point is the mean of five \SI{5}{s} loading trials. Error bars denote one standard error.
Solid curves are a joint zero-baseline Voigt fit to all four
resonances, with one shared Gaussian FWHM and independent
Lorentzian widths, centers, and amplitudes.
The common detuning $\Delta_{554}$ is referenced to the fitted
$^{137}\mathrm{Ba}$ $F'=1/2\rightarrow F''=1/2$ peak.
Dashed lines combine the calculated $6s6d\,{}^3D_1$ offsets
with measured $6s6p\,{}^3P_1^o$ offsets.
One common frequency translation aligns the calculated
$^{137}\mathrm{Ba}$ $F'=1/2\rightarrow F''=1/2$ line with zero. The remaining theoretical lines are not individually adjusted.
For the even isotopes, the $J'=1\rightarrow J''=1$ lines are shown.
The axis break omits a frequency interval.
Curves outside the measured scans are fit extrapolations.
}
\label{fig:loading554}
\end{figure*}

We perform the experiment in a linear Paul trap with a
resistively heated oven of naturally abundant barium metal.
During loading, the trap is operated at reduced confinement,
with a radial secular frequency of
\SI{1.3}{MHz} and an axial secular
frequency of \SI{100}{kHz}. After loading, we ramp up the radial confinement
for routine experiments.
In our experimental setup, shown in Fig.~\ref{fig:scheme}(b), the neutral atoms from the oven intersect the \SI{791}{nm}, \SI{554}{nm},
and \SI{532}{nm} photoionization beams in the trapping region. During
photoionization, \SI{493}{nm} Doppler-cooling and
\SI{650}{nm} repumping light are tuned to the corresponding transitions
of the selected isotope, $^{137}\mathrm{Ba}^{+}$ or
$^{135}\mathrm{Ba}^{+}$. A \SI{614}{nm} deshelving
beam is also applied during loading to remove population from the
metastable $5D_{5/2}$ manifold that can be populated by off-resonant
spontaneous Raman scattering from the intense \SI{532}{nm} field
\cite{Boguslawski2023}.
In addition, a second
\SI{493}{nm} cooling beam detuned by \SI{-210}{MHz} is
turned on together with the photoionization beams. This far-red-detuned
``protection'' beam provides additional cooling for newly loaded or
temporarily hot ions with large Doppler shifts and improves robustness during capture and crystallization.
At the trap, the \SI{791}{nm} and \SI{554}{nm} beams have diameters of
approximately \SI{700}{\micro\meter} and \SI{150}{\micro\meter},
respectively. The elliptical global Raman/ionization beam of \SI{532}{nm} has measured $1/e^2$ waists
of approximately \SI{13}{\micro\meter} and \SI{110}{\micro\meter}.   The corresponding nominal laser powers used during loading are $P_{791}=\SI{300}{\micro W}$, $P_{554}=\SI{150}{\micro W}$, and
$P_{532}=\SI{300}{mW}$. We image the fluorescence of the trapped ions using a qCMOS camera and a home-built $\mathrm{NA}=0.55$
objective. We check for dark ions embedded in the Coulomb crystals by inspecting the fluorescence images for localized gaps or defects in the otherwise regular ion-chain structure. Fig.~\ref{fig:scheme}(c) shows an exemplary long chain from our loading scheme.

We first characterize the loading spectrum by varying the frequency of the \SI{791}{nm} laser and focus on $^{137}\mathrm{Ba}^{+}$. During the measurements, we preheat the oven for \SI{60}{s} and then apply the ionization beams for \SI{10}{s}. The \SI{554}{nm} laser is fixed on the $F'=1/2\rightarrow F''=1/2$ resonance, and the \SI{532}{nm} beam is held at \SI{300}{mW}. The results are shown in Fig.~\ref{fig:loading791}.

To understand the result, we use a thermal-beam
excitation model to fit the observed profile, which incorporates the local slit acceptance, residual
Doppler shifts, and the spatial profile of the laser intensity. The signed
departure from perpendicular incidence is defined by
$\hat{\boldsymbol k}\cdot\hat{\boldsymbol z}_{a}=\sin\theta$,
where $\hat{\boldsymbol k}$ and $\hat{\boldsymbol z}_{a}$ denote
the laser and mean atomic-beam directions, respectively.
At an estimated oven temperature of \SI{500}{K}, the model gives
the observed width and approximate asymmetry with
$\theta=-5.1(2)^\circ$, a FWHM of $36(1)~\mathrm{MHz}$, and a peak rate of $5.2(3)~\mathrm{ions/s}$
(Supplemental Material \cite{SM}). Unless otherwise stated, uncertainties in parentheses refer to	the last quoted digits.
All 997 ions counted during the scan were identified as $^{137}\mathrm{Ba}^{+}$ using isotope-selective cooling and repumping, with no "dark" ions observed in the crystallized samples.
This distinction of "dark" ion is demonstrated in loading using \SI{413}{nm} two-photon loading scheme that yields substantially poorer isotope selectivity \cite{SM}.

\begin{figure}[!t]
\centering
\includegraphics[width=\columnwidth]{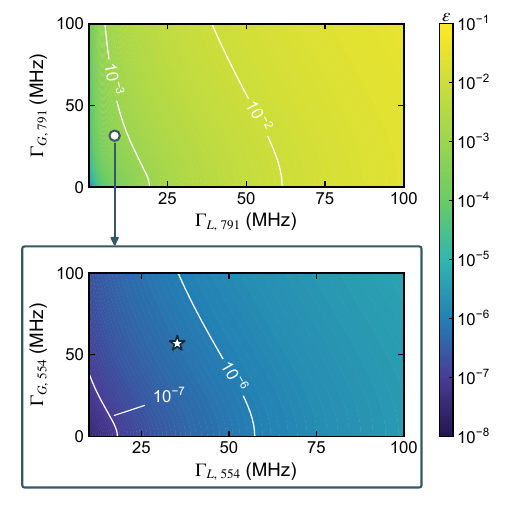}
\caption{Relayed isotope selectivity. Calculated undesired-isotope
excitation fractions after the \SI{791}{nm} step (upper map,
$\epsilon_1$) and after both resonant steps (lower map,
$\epsilon_2$), using a common logarithmic color scale.
The \SI{791}{nm} laser addresses the $^{137}\mathrm{Ba}$ $F'=1/2$ line.
The circle marks $(\Gamma_{L,791},\Gamma_{G,791})=(8.24,31.4)$ MHz
from the model analysis of the measured \SI{791}{nm} spectrum
in Fig.~\ref{fig:loading791} (see Supplemental Material \cite{SM}).
The arrow and frame indicate that the lower map retains the
isotope and intermediate-state weights at this circle while
varying the second-step widths. The \SI{554}{nm} laser
addresses the $^{137}\mathrm{Ba}$ $F'=1/2\rightarrow F''=1/2$ line.
The star marks $(\Gamma_{L,554},\Gamma_{G,554})=(35.2,56.9)$ MHz
from the joint fit in Fig.~\ref{fig:loading554}, giving
$\epsilon_2\simeq6.8\times10^{-7}$.
The calculation uses the line positions listed in the
Supplemental Material \cite{SM} and common widths for all
second-step components.}
\label{fig:relayedselectivity}
\end{figure}

To quantify first-step isotope discrimination, we model the first-step spectral response of isotope $A$
using transition strengths averaged over magnetic sublevels:
\begin{equation}
R_A^{(1)}(\nu_1)
=a_A\sum_{F'}s_{A,F'}\,
\mathcal V_1(\nu_1-\nu^{(1)}_{A,F'}),
\label{eq:firststepresponse}
\end{equation}
with $a_A$ the natural abundance, $\nu_1$ the
\SI{791}{nm} laser frequency, and $\nu^{(1)}_{A,F'}$ the resonance frequencies listed in the Supplemental Material \cite{SM}.
For $J=0\rightarrow J'=1$ excitation,
$s_{A,F'}=(2F'+1)/[3(2I_A+1)]$, where $I_A$ is the nuclear
spin. The peak-normalized Voigt profile $\mathcal V_1$ has
Lorentzian and Gaussian FWHM values $\Gamma_{L,791}$ and
$\Gamma_{G,791}$, respectively. We define the first-step
undesired isotope fraction as
\begin{equation}
\epsilon_1(\nu_1)
=\frac{\sum_{A\ne137}R_A^{(1)}(\nu_1)}
       {\sum_A R_A^{(1)}(\nu_1)}.
\label{eq:firststeperror}
\end{equation}
Based on our measurement in Fig.~\ref{fig:loading791}, we have $\Gamma_{G,791}=\SI{31.4}{MHz}$ and the nominal on-axis
power-broadened Lorentzian width $\Gamma_{L,791}=\SI{8.24}{MHz}$. At the selected $^{137}\mathrm{Ba}$ $F'=1/2$ line center, this model gives an undesired-isotope excitation fraction of
$\epsilon_1=5.94\times10^{-4}$
(Fig.~\ref{fig:relayedselectivity}), which is mostly affected the nearest competing line of $F'=1/2$ from $^{135}\mathrm{Ba}$ (\SI{357.4}{MHz} away).
A similar analysis for $^{135}\mathrm{Ba}$ with the same linewidth gives a calculated undesired-isotope
fraction of $1.56\times10^{-3}$.
We also present the corresponding comparisons with \SI{554}{nm} and
\SI{413}{nm} first-step excitation for both
isotopes in the Supplemental Material \cite{SM}.

With the \SI{791}{nm} laser fixed on the selected
$F'=1/2$ resonance, we scan the \SI{554}{nm} laser
across the well-separated loading peaks
(Fig.~\ref{fig:loading554}). Here, we apply the ionization beams for \SI{5}{s}. From the $F'=1/2$ intermediate state, electric-dipole
transitions can reach $F''=1/2$ and $3/2$ in
$6s6d\,{}^3D_1$ state. We first measure the loading spectrum for $^{137}\mathrm{Ba}^{+}$ and later change the frequencies of the \SI{791}{nm}, cooling, and repumping light for $^{135}\mathrm{Ba}^{+}$.
All four peaks are fitted jointly with zero-baseline Voigt
profiles sharing one Gaussian FWHM with independent
Lorentzian widths, centers, and amplitudes. The corresponding $^{135}\mathrm{Ba}$ resonances lie below the
$^{137}\mathrm{Ba}$ resonances by $595(2)$ and
$447(3)~\mathrm{MHz}$ for the
$F'=1/2\rightarrow F''=1/2$ and
$F'=1/2\rightarrow F''=3/2$, respectively,
which matches very well with the calculated separations of \SI{592.5}{MHz}
and \SI{450.7}{MHz}.
This agreement validates our later analysis on isotope selectivity for $^{133}\mathrm{Ba}$ \cite{SM}.
For the $F'=1/2\rightarrow F''=1/2$ and $F'=1/2\rightarrow F''=3/2$ branches,
the fitted peak loading rates are $4.8(3)$ and
$8.5(4)~\mathrm{ions/s}$ for $^{137}\mathrm{Ba}^{+}$,
and $2.6(2)$ and $3.4(2)~\mathrm{ions/s}$ for
$^{135}\mathrm{Ba}^{+}$, respectively.
The lower $^{135}\mathrm{Ba}^{+}$ loading rates may partly
reflect its lower natural abundance and less optimized cooling
and repumping conditions. Across the $^{137}\mathrm{Ba}$ ($^{135}\mathrm{Ba}$) \SI{554}{nm} scans, 1837 (668) ions
were identified as $^{137}\mathrm{Ba}^{+}$ ($^{135}\mathrm{Ba}^{+}$) by isotope-selective
fluorescence detection, with no undesired dark ions observed in the
crystallized samples.

The separated second-step lines can further suppress undesired-isotope
excitation remaining after the first step.
We retain each first-step contribution as
$p_{A,F'}=a_A s_{A,F'}
\mathcal V_1(\nu_1-\nu^{(1)}_{A,F'})$
and calculate the two-step response as
\begin{equation}
R_A^{(2)}(\nu_1,\nu_2)
=\sum_{F'}p_{A,F'}\sum_{F''}t_{A,F'F''}\,
\mathcal V_2(\nu_2-\nu^{(2)}_{A,F'F''}),
\label{eq:relay}
\end{equation}
with $\nu_2$ the \SI{554}{nm} laser frequency,
$\nu^{(2)}_{A,F'F''}$ the corresponding resonant frequencies,
$t_{A,F'F''}$ its normalized relative line strength,
and $\mathcal V_2$ the peak-normalized Voigt profile.
The sums include all allowed hyperfine paths
(see Supplemental Material \cite{SM}).
The undesired-isotope fraction after both steps is then
\begin{equation}
\epsilon_2(\nu_1,\nu_2)
=\frac{\sum_{A\ne137}R_A^{(2)}(\nu_1,\nu_2)}
       {\sum_A R_A^{(2)}(\nu_1,\nu_2)}.
\label{eq:relayerror}
\end{equation}
Note this definition preserves the isotope and intermediate-state
weights established by the first step and $\epsilon_2$ is therefore not a simple product of two independently normalized error fractions.

Figure~\ref{fig:relayedselectivity} summarizes the relayed
selectivity for $^{137}\mathrm{Ba}$. The upper map shows $\epsilon_1$
versus the first-step Lorentzian and Gaussian widths at measured isotope shifts. The lower
map fixes this response at the circle and varies the second-step
widths, with the \SI{554}{nm} laser on the
$^{137}\mathrm{Ba}$ $F'=1/2\rightarrow F''=1/2$ line center.
Using the fitted widths $(\Gamma_{L,554},\Gamma_{G,554})=(35.2,56.9)$ MHz
for the second step gives
$\epsilon_2\simeq6.8\times10^{-7}$, about an 870-fold
reduction from $\epsilon_1$, marked by the star.
The spectral-overlap model predicts substantial additional
suppression of undesired-isotope excitation through the
second resonant step. Experimentally, all 2834 ions counted
across the $^{137}\mathrm{Ba}^{+}$ spectroscopy scans were
identified as $^{137}\mathrm{Ba}^{+}$, with no dark ions
observed in the crystallized samples. Applying the same linewidth benchmark to $^{135}\mathrm{Ba}$ predicts
an approximately 290-fold reduction in undesired-isotope
excitation after the second step.
For a synthetic $^{133}\mathrm{Ba}$ source, the calculated second-step
separation of approximately \SI{5.0}{GHz} from the dominant
$^{132}\mathrm{Ba}$ component also provides strong additional spectral
discrimination \cite{SM}.

Beyond extremely high selectivity in our loading scheme, reusing the strong Raman laser for ionization greatly enhances the loading rate. We observe a linear dependence of the loading rate on the ionization \SI{532}{nm} intensity within the range of our scan \cite{SM}. A higher loading rate can be achieved by further increasing the \SI{532}{nm} laser power. Improving the final-step ionization efficiency allows rapid loading at reduced neutral atom flux, which reduces the detrimental charge exchange between ion and neutral atom.

In summary, we have demonstrated a three-step, ultraviolet-free
scheme for isotope-selective loading from natural-abundance
barium. The relayed selectivity from the first two resonant steps provides prominent loading purity ($\epsilon_2\simeq6.8\times10^{-7}$) and reusing of the strong Raman laser enhances loading rate (8.5(4) ions/s). Thus we establish a hardware-efficient route to larger barium-ion registers and motivate extension to source with higher temperature from laser ablation or synthetic source of the rare $^{133}\mathrm{Ba}$
\cite{SM}.

\begin{acknowledgments}
We acknowledge helpful discussions with Yao Lu and Jinming Cui. This project is supported by the Shanghai Municipal Science and Technology Co-research Program under Grant No. 25LZ2601000, the Quantum Science and Technology–National Science and Technology Major Project under Grants No. 2023ZD0300900, No. 2025ZD0300100 and No. 2025ZD0300101, and Scientific Research Innovation Capability Support Project for Young Faculty under Grant No. ZYGXQNJSKYCXNLZCXM-I14. Yanxin Liu is also supported by the China Postdoctoral Science Foundation under Grant No. 2025M783368.

\end{acknowledgments}

\makeatletter
\renewcommand{\bibsection}{%
  \par
  \begingroup
    \baselineskip26\p@
    \bib@device{\hsize}{72\p@}%
  \endgroup
  \nobreak\@nobreaktrue
  \addvspace{19\p@}%
}
\makeatother

\bibliographystyle{apsrev4-2}
\IfFileExists{arxiv/ref.bib}
  {\bibliography{arxiv/ref}}
  {\bibliography{ref}}
\clearpage
\makesupplementtitle

\IfFileExists{arxiv/supplement_body.tex}
  {\raggedbottom
\setcounter{section}{0}
\setcounter{table}{0}
\setcounter{figure}{0}
\setcounter{equation}{0}
\renewcommand{\thesection}{\Roman{section}}
\renewcommand{\thetable}{S\arabic{table}}
\renewcommand{\thefigure}{S\arabic{figure}}
\renewcommand{\theequation}{S\arabic{equation}}
\renewcommand{\theHsection}{supp.\arabic{section}}
\renewcommand{\theHtable}{supp.\arabic{table}}
\renewcommand{\theHfigure}{supp.\arabic{figure}}
\renewcommand{\theHequation}{supp.\arabic{equation}}

\section{Review of photoionization pathways}
\label{sec:suppathways}

\begin{figure}[h]
\centering
\includegraphics[width=\columnwidth]{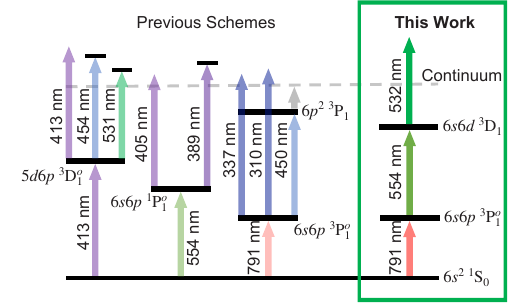}
\caption{Representative Ba photoionization pathways.
Here the energy spacings are
schematic and not to scale.}
\label{fig:suppathways}
\end{figure}

We summarize previously reported photoionization schemes in Fig.~\ref{fig:suppathways} in comparison to our three-photon scheme, which is
highlighted by the green outline. Two-photon photoionization schemes using two \SI{413}{nm} photons, one \SI{413}{nm} photon and one \SI{454}{nm} photon,  and
one \SI{413}{nm} photon and one \SI{531}{nm} photon are discussed in Refs.~\cite{Leschhorn2012,Choi2024,Wall2026}. Other two-photon schemes use the strong \SI{554}{nm}
transition for excitation, followed by nonresonant
photoionization at \SI{405}{nm} or resonantly enhanced
ionization near \SI{389}{nm}
\cite{Leschhorn2012,White2022,Greenberg2024}.
The highly isotope selective approach using \SI{791}{nm} laser is first demonstrated in Refs.~\cite{Steele2007,Wang2011} followed by a UV photon of \SI{337}{nm} or \SI{310}{nm} for ionization. Furthermore, another three-photon pathway is discussed in Refs.~\cite{Graham2014,Lewty2014}. Here a \SI{791}{nm} photon is followed by
a \SI{450}{nm} photon to resonantly excite the atom to the $6p^2\,{}^3P_1$ state. Final ionization can be achieved by the absorption of a
third photon.

\section{Model for the \SI{791}{nm} loading spectrum}
\label{sec:sup791}

We fit the measured \SI{791}{nm} loading spectrum using
a power-broadened two-level response averaged over
position, atomic speed, and the angular spread in the
laser--atomic-beam plane. The fit determines the peak
rate, spectral width, and effective departure from
perpendicular incidence, and supplies the Doppler widths
for the selectivity comparison in
Sec.~\ref{sec:supcomparison}.

First, we describe the local excitation by the steady-state
response of a closed two-level atom,
\begin{equation}
L(\Delta;\boldsymbol r,v,\alpha)=
\frac{s(\boldsymbol r)}{1+s(\boldsymbol r)
+4[\Delta-q_D(v,\alpha;\theta)]^2/\Gamma_0^2}.
\label{eq:supExc}
\end{equation}
Here $\Delta=\nu_L-\nu_0$ is the laser detuning from the
zero-velocity resonance, $s(\boldsymbol r)=
I(\boldsymbol r)/I_{\rm sat}$ is the local saturation
parameter, and $\Gamma_0$ is the natural FWHM.
To first order in the angular spread, the Doppler shift is
$q_D\simeq v(\sin\theta+\alpha\cos\theta)/\lambda_{791}$,
where $v$ is the atomic speed, $\lambda_{791}$ is the laser wavelength,
and $\theta$ is the signed departure from perpendicular
incidence. The angle $\alpha$ describes the trajectory
projected onto the plane containing the mean atomic-beam
and laser axes, measured from the mean atomic-beam direction.
Positive $\theta$ and $\alpha$ increase the Doppler shift
near perpendicular incidence, as illustrated in
Fig.~\ref{fig:sup791geometry}.

\begin{figure}[t]
\centering
\includegraphics[width=\columnwidth]{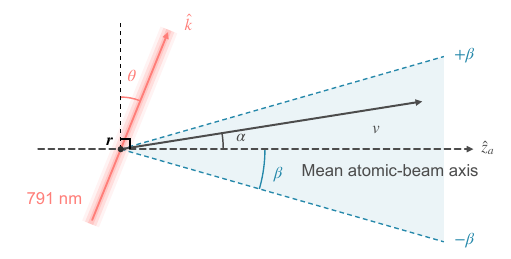}
\caption{Angular geometry of the
\SI{791}{nm} model in the laser--atomic-beam plane.
The horizontal dashed arrow marks the mean atomic-beam
direction $\hat{z}_a$.
The laser direction $\hat{k}$ departs from the perpendicular
by $\theta$. At position $\boldsymbol r$, $\alpha$ is the
angle between the projected atomic velocity and $\hat{z}_a$.
The angular integral averages this single angle over the
shaded sector $-\beta\leq\alpha\leq\beta$, where $\beta$
is the acceptance half-angle.
Positive $\theta$ is shown, and angles are exaggerated
for clarity.}
\label{fig:sup791geometry}
\end{figure}

Averaging over the modeled interaction volume $\Omega$, atomic
speeds, and accepted trajectory angles gives
\begin{equation}
\begin{aligned}
\mathcal S_{791}(\Delta;\theta)={}&
\int_{\Omega}d^3r\,W(\boldsymbol r)
\int_0^\infty dv\,f_n(v)\\
&\times\int_{-\beta}^{\beta}\frac{d\alpha}{2\beta}
L(\Delta;\boldsymbol r,v,\alpha).
\end{aligned}
\label{eq:supSpectrum}
\end{equation}
The spatial integral averages over three-dimensional
positions, while the angular integral averages the single
projected angle $\alpha$ within the laser--atomic-beam plane.
Out-of-plane angular deviations enter the Doppler projection
only at second order in this paraxial approximation.
The spatial density weight $W(\boldsymbol r)$ and local speed distribution
$f_n(v)$ are normalized and treated as independent of each
other and of the uniform angular distribution on
$[-\beta,\beta]$. Thus $\mathcal S_{791}$ is a dimensionless
excitation response, equal to twice the average steady-state
excited-state population in the two-level model.
We relate it to the measured loading rate through
\begin{equation}
R_{\rm fit}(\nu_L)
=C\,\mathcal S_{791}(\nu_L-\nu_0;\theta)+R_{\rm bg}.
\label{eq:supFit}
\end{equation}
Here $C$ is a rate scale and $R_{\rm bg}$ is a constant
background, both nonnegative and in units of ions/s.
The fit varies $C$, $R_{\rm bg}$, $\nu_0$, and $\theta$.
The scale $C$ incorporates the atom number and the
efficiency of subsequent excitation, ionization, and capture,
taken to be independent of velocity over the contributing ensemble.

Second, we fix the laser and oven parameters entering
this response. The natural width is
$\Gamma_0=\SI{0.118}{MHz}$ \cite{Scielzo2006}.
For a Gaussian beam,
$s(\boldsymbol r)=s_0e^{-2r_\perp^2/w_{791}^2}$, where
$r_\perp$ is the distance from the laser axis and
$w_{791}$ is the intensity $1/e^2$ radius.
With $P_{791}=\SI{300}{\micro W}$,
$w_{791}=\SI{350}{\micro\meter}$, and the effective
saturation intensity $I_{\rm sat}=\SI{0.32}{W/m^2}$,
we obtain $s_0=2P_{791}/(\pi w_{791}^2I_{\rm sat})
=4.9\times10^3$ and a beam-center Lorentzian FWHM
$\Gamma_0\sqrt{1+s_0}=\SI{8.24}{MHz}$.
The spatial average retains the local intensity variation.

At the adopted oven temperature $T=\SI{500}{K}$,
we use $f_n(v)\propto v^2e^{-mv^2/(2k_BT)}$, with
$m=137\,\mathrm u$ and Boltzmann's constant $k_B$.
This density distribution follows from weighting the
effusive flux distribution by the residence time,
which is proportional to $1/v$.
The \SI{300}{\micro\meter}-wide slit, \SI{34}{mm}
from the trap center, sets the nominal angular half-width
$\beta=\tan^{-1}(0.15/34)=\SI{0.253}{\degree}$.
For the spatial average, we approximate the effective
loading region $\Omega$ by a cube of side
\SI{200}{\micro\meter} centered at the laser intersection.
This window represents spatial selection by subsequent
excitation, ionization, and capture, with uniform acceptance
inside the cube. Within $\Omega$, we use the normalized
neutral-density weight
$W(\boldsymbol r)\propto e^{-2r_{a,\perp}^2/r_a^2}$,
where $r_{a,\perp}$ is the distance from the atomic-beam
axis and $r_a=\SI{1}{mm}$ is the adopted neutral-beam
density $1/e^2$ radius.

With these inputs fixed, an inverse-variance-weighted
least-squares fit to eight frequency points, each averaged
over five trials, gives $\theta=-5.1(2)^\circ$,
a FWHM of $36(1)~\mathrm{MHz}$, and a peak loading
rate above background of $5.2(3)~\mathrm{ions/s}$.
The best-fit background is zero.
The uncertainties are profile intervals at
$\Delta\chi^2=1$ with the model inputs held fixed.
Thus $\theta$ is the effective angle for the adopted
source model. The plotted detuning is referenced
to the fitted spectral maximum.

Finally, we obtain the Voigt Gaussian widths by matching
the variance of the projected velocity
$u\simeq v(\sin\theta+\alpha\cos\theta)$ to a Gaussian
profile. For the independent speed and angular
distributions used above,
\begin{equation}
\begin{aligned}
\sigma_u^2={\rm Var}(u)\simeq{}&
{\rm Var}(v)\sin^2\theta
+\langle v^2\rangle\frac{\beta^2}{3}\cos^2\theta,\\
\Gamma_{G,i}={}&
\frac{2\sqrt{2\ln2}\,\sigma_u}{\lambda_i}.
\end{aligned}
\label{eq:supVelocityWidth}
\end{equation}
For the adopted speed distribution,
$\langle v^2\rangle=3k_BT/m$ and
$\operatorname{Var}(v)=(3-8/\pi)k_BT/m$.
The unrounded fitted angle gives a
Gaussian-equivalent velocity FWHM of $\Delta v_{\mathrm{eff}}^{\mathrm{FWHM}}=\SI{24.9}{m/s}$,
corresponding to $\Gamma_{G,i}=(31.4,44.9,60.2)$ MHz
for $\lambda_i=(791,554,413)$ nm.
The widths depend on $|\theta|$, so the selectivity
comparison uses nonnegative angles.
For the first-step selectivity comparison, we combine these
Gaussian widths with Lorentzian FWHM values of
\SI{8.24}{MHz} at \SI{791}{nm},
\SI{19.0}{MHz} at \SI{554}{nm}, and
\SI{9.15}{MHz} at \SI{413}{nm}.
The latter two are the natural widths \cite{Niggli1987}.
Together with the line positions in
Table~\ref{tab:sup137offsets}, these widths define
the Voigt profiles used in the comparison.

\begin{table*}[!t]
\centering
{\suppTableStyle
\setlength{\tabcolsep}{1.5pt}
\begin{tabular*}{\textwidth}{@{\extracolsep{\fill}}lccccccccc@{}}
\toprule
& \multicolumn{9}{c}{$\nu_{A,F'}-\nu_{138}$ (MHz)} \\
\cmidrule(lr){2-10}
First step & $^{138}\mathrm{Ba}$ & \multicolumn{3}{c}{$^{137}\mathrm{Ba}$}
& $^{136}\mathrm{Ba}$ & \multicolumn{3}{c}{$^{135}\mathrm{Ba}$} & $^{134}\mathrm{Ba}$ \\
\cmidrule(lr){3-5}\cmidrule(lr){7-9}
& -- & $F'=5/2$ & $F'=3/2$ & $F'=1/2$ & --
& $F'=5/2$ & $F'=3/2$ & $F'=1/2$ & -- \\
\midrule
\SI{791}{nm} & 0 & 1898(2) & -924.5(7)
& \textbf{-2743(2)} & 108.5(3) & 1754(2)
& -782(1) & \textbf{-2386(2)} & 122(3) \\
\SI{554}{nm} & 0 & 63.4(3) & \textbf{274.56(9)}
& $549.5(1)^{\dagger}$ & 128.0(4) & 120.6(3)
& \textbf{323.4(2)} & $549.5(1)^{\dagger}$ & 143(1) \\
\SI{413}{nm} & 0 & \textbf{651.4} & 223.5 & -43.7
& 178.1 & \textbf{633.3} & 248.8 & 10.9 & 124.2 \\
\bottomrule
\end{tabular*}}
\caption{First-step isotope and hyperfine peak offsets relative to
$^{138}\mathrm{Ba}$ for the natural-Ba spectral comparison.
The \SI{791}{nm} entries are from Ref.~\cite{Yu2018},
with the $^{134}\mathrm{Ba}$ value from
Ref.~\cite{Grundevik1983}.
The \SI{554}{nm} entries are from
Ref.~\cite{VanWijngaarden1995},
with the $^{134}\mathrm{Ba}$ value from Ref.~\cite{Baird1979}.
The \SI{413}{nm} entries reproduce the peak list in Table~III
of Ref.~\cite{Choi2024}.
Parentheses give the published $1\sigma$ uncertainties,
rounded to one significant digit and referring to the last
quoted digit.
Calculations use the unrounded published values.
Individual uncertainties are not given in the published
\SI{413}{nm} peak list.
The dagger marks the unresolved $^{135,137}\mathrm{Ba}$
feature whose frequency is used for both component positions.
Bold entries identify the components selected for the
$^{137}\mathrm{Ba}$ and $^{135}\mathrm{Ba}$
first-step selectivity comparisons.
Here, \SI{554}{nm} refers to the
$6s^2\,{}^1S_0\rightarrow6s6p\,{}^1P_1^o$
first-step transition.}
\label{tab:sup137offsets}
\end{table*}

\section{Comparison with the \SI{413}{nm} loading scheme}
\label{sec:sup413loading}

We also measured the loading spectrum with the two-photon ionization regime using \SI{413}{nm} laser. This measurement provides an
experimental comparison with the three-step scheme.
The excitation frequency was scanned around the
$^{137}\mathrm{Ba}$
$6s^2\,{}^1S_0\rightarrow
5d6p\,{}^3D_1^o(F'=5/2)$ resonance.
Two \SI{413}{nm} frequency components were used:
a weak excitation component with a power of
\SI{5}{\micro\watt}, chosen to limit power broadening,
and a stronger component with a power of
\SI{330}{\micro\watt}, blue-detuned by approximately
\SI{440}{MHz} from the excitation component, to provide
the ionizing photon.
The beam waist was approximately \SI{80}{\micro\meter},
quoted as the intensity $1/e^2$ radius.
The laser incidence angle relative to the atomic beam was
carefully adjusted to reduce Doppler broadening.

Figure~\ref{fig:sup413loading} shows loading spectrum of both $^{135}\mathrm{Ba}^{+}$ and $^{137}\mathrm{Ba}^{+}$ as we scan the detuning of \SI{413}{nm} laser. We also show an example of mixed loading of both $^{135}\mathrm{Ba}^{+}$ and $^{137}\mathrm{Ba}^{+}$ with low selectivity.
The nearest competing $^{135}\mathrm{Ba}$ $F'=5/2$
resonance lies only \SI{18.1}{MHz} below the selected line
\cite{Choi2024}, with a natural linewidth of \SI{9.15}{MHz}. This combination gives significant overlap between the two isotopes and hence low loading selectivity.

Besides isotope discrimination, we observe that a high-power \SI{532}{nm} Raman beam can assist
ionization from the $5d6p\,{}^3D_1^o$ state.
Resonantly enhanced ionization from this state has also
been demonstrated using the broad \SI{530.8}{nm}
autoionizing resonance \cite{Wall2026}.
These approaches enhance final-step ionization rather
than resolve the overlap between the closely spaced
\SI{413}{nm} excitation resonances.
The present scheme addresses isotope discrimination
through two complementary spectral features: the larger
first-step separation from competing isotopes and an
additional narrowband isotope filter provided by the
resonant \SI{554}{nm} step.

\begin{figure}[t]
    \centering
    \includegraphics[width=\columnwidth]{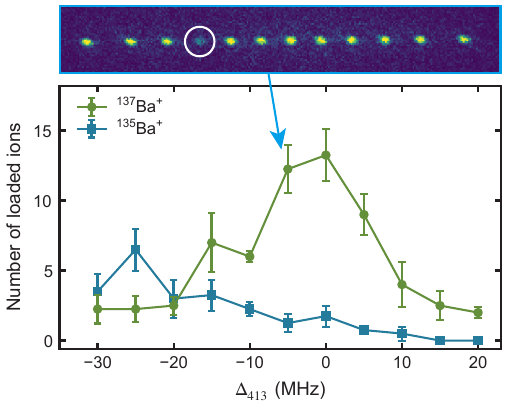}
\caption{
Mean numbers of $^{137}\mathrm{Ba}^{+}$ (green circles)
and $^{135}\mathrm{Ba}^{+}$ (blue squares) ions after
\SI{300}{s} of loading, measured while scanning the weak
\SI{413}{nm} excitation component across the
$^{137}\mathrm{Ba}$ $F'=5/2$ resonance.
Each point averages four trials.
Error bars denote one standard error of the mean.
Connecting lines are guides to the eye.
The detuning $\Delta_{413}$ is referenced to the frequency
yielding the maximum measured mean $^{137}\mathrm{Ba}^{+}$ count.
The top image shows an exemplary experiment shot at
$\Delta_{413}=-\SI{5}{MHz}$ (arrow), containing
11 $^{137}\mathrm{Ba}^{+}$ ions (bright) and one
$^{135}\mathrm{Ba}^{+}$ ion (dim, marked by the white circle).
Two additional frequency components in the \SI{493}{nm}
light addressed $^{135}\mathrm{Ba}^{+}$, allowing both
isotopes to fluoresce.
}
    \label{fig:sup413loading}
\end{figure}

\section{King-plot estimates for barium-132 and barium-133}
\label{sec:supking}

This section provides the first-step resonance frequencies
needed for the $^{133}\mathrm{Ba}$ source comparison in
Sec.~\ref{sec:supcomparison}. King plots relate the
neutral-Ba centroid isotope shifts to the measured
\SI{493}{nm} $\mathrm{Ba}^+$ shifts. The predicted
centroids, combined with scaled hyperfine splittings,
give the relevant $^{132}\mathrm{Ba}$ and
$^{133}\mathrm{Ba}$ line positions for the
\SI{791}{nm}, \SI{554}{nm}, and \SI{413}{nm} transitions.

First, we calibrate the centroid-shift relations using
the measured $A=134$--137 isotopes.
Here $A$ is the isotope mass number, and
$i\in\{791,554,413\}$ labels the neutral-Ba transitions
by their wavelengths in nanometers. The quantity
$\delta\nu_i^{A,138}$ denotes the hyperfine
center-of-gravity frequency of isotope $A$ minus the
$^{138}\mathrm{Ba}$ resonance frequency.
To first order in the hyperfine interaction, the centroid
for $^{135,137}\mathrm{Ba}$ is the degeneracy-weighted average
of the $F'=1/2,3/2,5/2$ component frequencies, with weights
$2F'+1=2,4,6$, respectively.
The calibration uses the unrounded published centroids
summarized in Table~\ref{tab:supKingInputs}:
Refs.~\cite{Yu2018,Grundevik1983} for \SI{791}{nm},
Refs.~\cite{VanWijngaarden1995,Baird1979} for
\SI{554}{nm}, and Ref.~\cite{Choi2024} for
\SI{413}{nm}.

At leading order, the common nuclear mass and charge-radius dependence
of the isotope shifts gives a linear King relation
between the neutral and ionic transitions.
Using isotope masses approximated by $A\,\mathrm u$,
we define the modified shifts
$\widetilde{\delta\nu}_i^{A,138}
=\mu_{A,138}\delta\nu_i^{A,138}$, with
$\mu_{A,138}=138A\,\mathrm u/(138-A)$ and
$\mathrm u$ the atomic mass constant. The relation is
\begin{equation}
\widetilde{\delta\nu}_i^{A,138}
=a_i\widetilde{\delta\nu}_{493}^{A,138}+b_i,
\label{eq:supKing}
\end{equation}
where $a_i$ and $b_i$ are the fitted slope and intercept.
The reference is the $\mathrm{Ba}^{+}$
$6s\,{}^2S_{1/2}\rightarrow6p\,{}^2P_{1/2}$ transition.
Its centroid shifts for $A=(134,135,136,137)$ are
$(223.2(2),350.0(2),180.2(2),272.2(2))$ MHz
\cite{Imgram2019}. The fits account for independent
Gaussian measurement uncertainties in both axes.
Figure~\ref{fig:supkingplots} shows the fitted lines
and residuals, and Table~\ref{tab:supKingFits} lists
the coefficients. For uncertainty propagation, we multiply the covariance
matrix of the fitted slope $a_i$ and intercept $b_i$
by the reduced chi-squared $\chi_i^2/2$ when this value
exceeds unity, to account for excess scatter in the
calibration data. Otherwise, we retain the original
covariance matrix.

The fitted relations then give the target centroid shifts as
\begin{equation}
\delta\nu_{i,\mathrm{pred}}^{A,138}
=a_i\delta\nu_{493}^{A,138}
+\frac{b_i}{\mu_{A,138}}.
\label{eq:supKingPrediction}
\end{equation}
For $^{132}\mathrm{Ba}$ and $^{133}\mathrm{Ba}$,
the reference inputs are
$\delta\nu_{493}^{132,138}=278.8(2)~\mathrm{MHz}$
\cite{Imgram2019} and
$\delta\nu_{493}^{133,138}=373(4)~\mathrm{MHz}$
\cite{Hucul2017}, respectively.
In the modified-shift coordinates, the $^{133}\mathrm{Ba}$
input lies within the calibration range, while the
$^{132}\mathrm{Ba}$ input lies just below it.

Second, we obtain the $^{133}\mathrm{Ba}$ hyperfine
components by adding their shifts to the predicted
centroids. For $I=1/2$ and $J'=1$, the excited levels
are $F'=1/2$ and $3/2$, with no electric-quadrupole
term. For each transition, $A_{\mathrm{hfs}}$ denotes
the magnetic-dipole hyperfine constant of its
$^{133}\mathrm{Ba}$ upper state, expressed in MHz.
The $J=0$ ground state has no first-order hyperfine
splitting, so the component positions are
\begin{equation}
\begin{aligned}
\delta\nu_{i,1/2}^{133,138}
&=\delta\nu_{i,\mathrm{pred}}^{133,138}-A_{\mathrm{hfs}},\\
\delta\nu_{i,3/2}^{133,138}
&=\delta\nu_{i,\mathrm{pred}}^{133,138}+\tfrac12 A_{\mathrm{hfs}}.
\end{aligned}
\label{eq:supHFS}
\end{equation}

These constants follow from nuclear-$g$-factor scaling
of the measured $^{137}\mathrm{Ba}$ constants, neglecting
differential hyperfine anomalies.
The scaling factor $\kappa=-2.470$ is taken from the
ratio of the $^{133}\mathrm{Ba}^{+}$ and
$^{137}\mathrm{Ba}^{+}$ $6s\,{}^2S_{1/2}$ hyperfine
constants in Ref.~\cite{Hucul2017}.
Its negative sign reverses the hyperfine ordering
\cite{Hoehle1976}.
For the \SI{791}{nm}, \SI{554}{nm}, and \SI{413}{nm}
upper states, the measured $^{137}\mathrm{Ba}$ constants
are $1150(1)$, $-109.5(1)$, and $172.9(3)$ MHz
\cite{Yu2018,VanWijngaarden1995,Choi2024}.
Multiplying their unrounded values by $\kappa$ gives
$A_{\mathrm{hfs}}$ values of $-2840.253$, $270.465$,
and $-427.063$ MHz for the three $^{133}\mathrm{Ba}$
states in the same order.

In summary, the King-plot centroids and scaled hyperfine
splittings give the first-step line positions in
Table~\ref{tab:sup133offsets}. The selected
$^{133}\mathrm{Ba}$ resonances are at
$3056(4)$, $392(3)$, and $738(7)$ MHz relative to
$^{138}\mathrm{Ba}$ for the \SI{791}{nm},
\SI{554}{nm}, and \SI{413}{nm} transitions, respectively.
The unrounded predictions provide the spectral inputs
for the source comparison in Sec.~\ref{sec:supcomparison}.

\begin{figure*}[!t]
\centering
\includegraphics[width=0.85\textwidth]{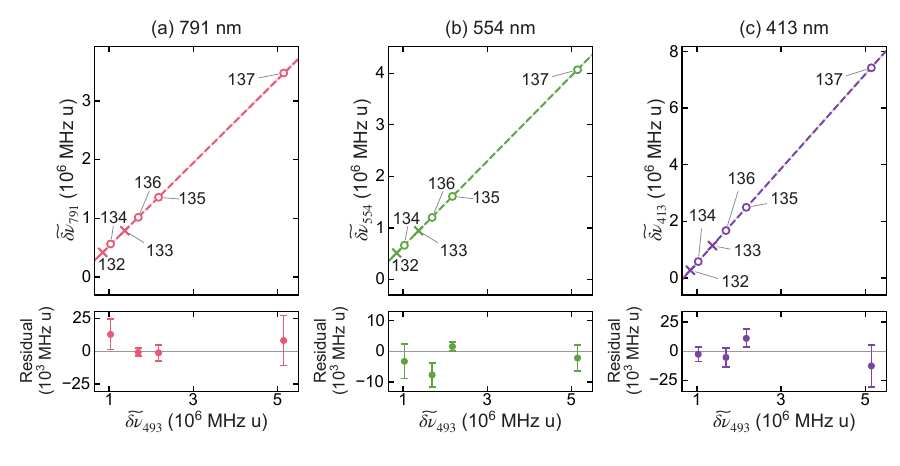}
\caption{King plots relating the modified neutral-Ba
first-step centroid isotope shifts to those of the
\SI{493}{nm} $\mathrm{Ba}^+$ reference transition.
Open circles show the measured centroids for $A=134$--137,
and crosses show the $^{132}\mathrm{Ba}$ and
$^{133}\mathrm{Ba}$ predictions.
Dashed lines are fits accounting for uncertainties
in both axes. Lower panels show vertical residuals,
with error bars combining the two-axis measurement
uncertainties in quadrature using the fitted slopes.}
\label{fig:supkingplots}
\end{figure*}

\begin{table}[!t]
\centering
{\suppTableStyle
\begin{tabular}{crrr}
\toprule
& \multicolumn{3}{c}{$\delta\nu_i^{A,138}$ (MHz)}\\
\cmidrule(lr){2-4}
$A$ & \SI{791}{nm} & \SI{554}{nm} & \SI{413}{nm}\\
\midrule
134 & 122(3) & 143(1) & 124(1)\\
135 & 219(1) & 259.3(2) & 401(1)\\
136 & 108.5(3) & 128.0(4) & 178.1(8)\\
137 & 184(1) & 215.2(2) & 392.9(9)\\
\bottomrule
\end{tabular}
}
\caption{Published neutral-Ba centroid isotope shifts
relative to $^{138}\mathrm{Ba}$ used for calibration.
Parentheses give $1\sigma$ uncertainties in the last
quoted digit, rounded to one significant digit.
The fits use the unrounded published inputs.}
\label{tab:supKingInputs}
\end{table}

\begin{table}[!t]
\centering
{\suppTableStyle
\begin{tabular}{crrr}
\toprule
$i$ (nm) & $a_i$ & $b_i$ ($10^6$ MHz\,u) & $\chi_i^2/2$\\
\midrule
791 & 0.707796 & $-0.177821$ & 0.77\\
554 & 0.827889 & $-0.190811$ & 2.80\\
413 & 1.668131 & $-1.144118$ & 1.56\\
\bottomrule
\end{tabular}
}
\caption{King-plot coefficients in Eq.~\eqref{eq:supKing}.
The slope $a_i$ is dimensionless, and $b_i$ is given
in units of $10^6$ MHz\,u.
The last column gives the reduced chi-squared
for two residual degrees of freedom.}
\label{tab:supKingFits}
\end{table}

\begin{table}[!t]
\centering
{\suppTableStyle
\setlength{\tabcolsep}{3pt}
\begin{tabular}{c c cc cc}
\toprule
& \multicolumn{5}{c}{Frequency offset from $^{138}\mathrm{Ba}$ (MHz)} \\
\cmidrule(lr){2-6}
First step & $^{138}\mathrm{Ba}$
& \multicolumn{2}{c}{$^{133}\mathrm{Ba}$}
& $^{132}\mathrm{Ba}$ & $^{134}\mathrm{Ba}$ \\
\cmidrule(lr){3-4}
& -- & $F'=1/2$ & $F'=3/2$ & -- & -- \\
\midrule
\SI{791}{nm} & 0 & \textbf{3056(4)} & -1205(3) & 139(2) & 122(3) \\
\SI{554}{nm} & 0 & -14(3) & \textbf{392(3)} & 168(1) & 143(1) \\
\SI{413}{nm} & 0 & \textbf{738(7)} & 97(7) & 88(2) & 124.2 \\
\bottomrule
\end{tabular}}
\caption{First-step frequency offsets relative to
$^{138}\mathrm{Ba}$ for the $^{133}\mathrm{Ba}$ source
comparison. The $^{132}\mathrm{Ba}$ and
$^{133}\mathrm{Ba}$ predictions follow
Eqs.~\eqref{eq:supKingPrediction} and \eqref{eq:supHFS}.
Parentheses give propagated $1\sigma$ uncertainties
in the last quoted digit, rounded to one significant digit.
For the predicted entries, uncertainties combine the
fitted-line covariance and independent target-reference
errors, together with measured hyperfine-constant errors
for $^{133}\mathrm{Ba}$, with $\kappa$ held fixed.
The $^{134}\mathrm{Ba}$ positions are from
Table~\ref{tab:sup137offsets}.
Bold entries mark the selected $^{133}\mathrm{Ba}$
components. Calculations use the unrounded predictions.}
\label{tab:sup133offsets}
\end{table}

\section{First-step selectivity for thermal and ablation sources}
\label{sec:supcomparison}

We compare the \SI{791}{nm}, \SI{554}{nm}, and \SI{413}{nm}
first-step transitions for two source conditions:
natural-abundance thermal barium and a synthetic
$^{133}\mathrm{Ba}$ ablation source.
Within each source model, the three transitions share the
same projected-velocity width.
Here \SI{554}{nm} denotes the conventional
$6s^2\,{}^1S_0\rightarrow6s6p\,{}^1P_1^o$ transition.

For both sources, the relative excitation
response of isotope $A$ is
\begin{equation}
R_A^{(1)}(\nu)
=a_A\sum_{F'}s_{A,F'}\,
\mathcal V\!\left(
\nu-\nu_{A,F'};\Gamma_L,\Gamma_G
\right),
\label{eq:rate}
\end{equation}
where $\nu$ is the laser frequency, $a_A$ is the source
abundance, and $\nu_{A,F'}$ is the component frequency.
Averaging over magnetic sublevels for
$J=0\rightarrow J'=1$ excitation gives
$s_{A,F'}=(2F'+1)/[3(2I_A+1)]$,
where $I_A$ is the nuclear spin. Each even isotope has one component with
unit strength.
The peak-normalized Voigt profile $\mathcal V$ has
Lorentzian FWHM $\Gamma_L$ and Gaussian FWHM
$\Gamma_G=\Delta v_{\mathrm{eff}}^{\mathrm{FWHM}}/\lambda$,
where $\Delta v_{\mathrm{eff}}^{\mathrm{FWHM}}$ is the FWHM
of the Gaussian profile representing the projected velocity
distribution and $\lambda$ is the transition wavelength.
All isotope and hyperfine components within a transition
share these widths.

For target isotope $A_t$, the excitation selectivity is
\begin{equation}
\eta_{A_t}(\nu)
=\frac{R_{A_t}^{(1)}(\nu)}{\sum_A R_A^{(1)}(\nu)},
\qquad A_t\in\{137,135,133\}.
\label{eq:selectivity}
\end{equation}
The responses include all hyperfine components of each
isotope in the source model.
For the \SI{791}{nm} response, this gives
$\eta_{137}=1-\epsilon_1$, consistent with the main text.
At each source's reference angle, we select the component
with the highest line-center selectivity for each transition
and target isotope.

\begin{figure}[t]
\centering
\includegraphics[width=\columnwidth]{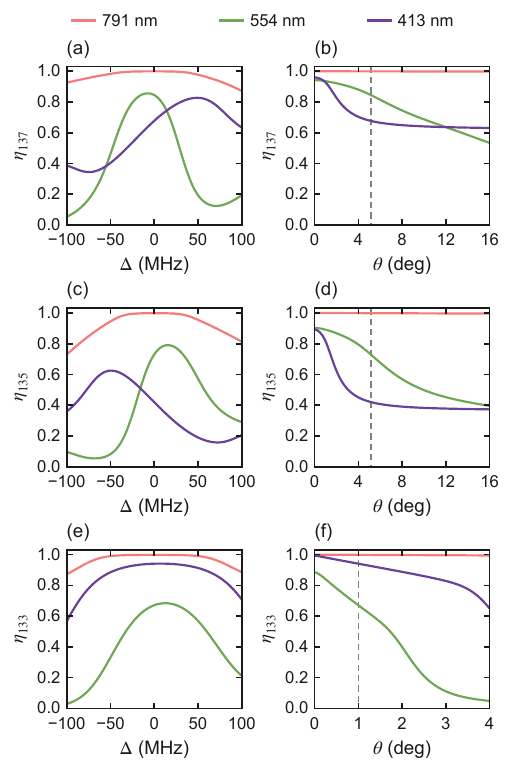}
\caption{Calculated first-step excitation selectivity
$\eta_{A_t}$ for (a,b) $^{137}\mathrm{Ba}$,
(c,d) $^{135}\mathrm{Ba}$, and (e,f) $^{133}\mathrm{Ba}$.
Colors distinguish the \SI{791}{nm}, \SI{554}{nm}, and
\SI{413}{nm} first-step transitions.
The left column shows detuning scans at the reference
angles of \SI{5.1}{\degree} for (a,c) and
\SI{1}{\degree} for (e).
The right column shows line-center selectivity versus angle,
with dashed lines marking these reference angles. The Doppler widths depend on $|\theta|$, so only
nonnegative angles are shown.
Panels (a--d) use natural abundances and the same thermal
velocity distribution.
Panels (e,f) use the synthetic-source conditions in the text.
For each transition, the selected hyperfine component
remains fixed during the angular scan.
The laser tracks its resonance after compensation of the
mean Doppler shift, and $\Delta=0$ denotes this resonance.}
\label{fig:selectivitycomparison}
\end{figure}

First, we evaluate $^{137}\mathrm{Ba}$ and $^{135}\mathrm{Ba}$
under the thermal-source conditions established in
Sec.~\ref{sec:sup791}.
For $A=134,135,136,137,138$, the abundance fractions are
$(0.02417,0.06592,0.07854,0.11232,0.71698)$,
respectively \cite{Meija2016}.
We use the unrounded component positions underlying
Table~\ref{tab:sup137offsets}.
We use the projected-velocity distribution and
transition-specific Voigt linewidths established in
Sec.~\ref{sec:sup791}.
The same parameters are used for both target isotopes.

For both $^{137}\mathrm{Ba}$ and $^{135}\mathrm{Ba}$,
the highest line-center selectivity at the reference angle
is obtained with $F'=1/2$ at \SI{791}{nm},
$F'=3/2$ at \SI{554}{nm}, and $F'=5/2$ at \SI{413}{nm}.
The calculated line-center selectivities are
$(99.94,84.52,67.72)\%$ for $^{137}\mathrm{Ba}$ and
$(99.84,72.93,42.03)\%$ for $^{135}\mathrm{Ba}$.
The \SI{791}{nm} transition combines a
\SI{357.4}{MHz} separation between the selected
$^{137}\mathrm{Ba}$ and $^{135}\mathrm{Ba}$ components
with smaller Doppler and Lorentzian widths.
It therefore gives the highest line-center selectivity
and the broadest high-selectivity region in detuning and
angle (Fig.~\ref{fig:selectivitycomparison}(a--d)).

Second, we apply the same comparison to microgram-scale
$\mathrm{BaCl}_2$ ablation sources for $^{133}\mathrm{Ba}^+$
\cite{White2022,GreenbergAblation2024}.
We adopt a composition of 2\% $^{133}\mathrm{Ba}$,
97\% $^{132}\mathrm{Ba}$, and 1\% $^{134}\mathrm{Ba}$
\cite{ChristensenThesis2020}.
The component positions are the unrounded King-plot and
hyperfine-scaling estimates underlying
Table~\ref{tab:sup133offsets}, with the measured
$^{134}\mathrm{Ba}$ position retained. Based on the neutral-Ba plume characterization in
Ref.~\cite{White2022}, we adopt a collimated-source
benchmark with longitudinal velocity FWHM
$\Delta v_{\parallel}^{\mathrm{FWHM}}=\SI{2.50}{km/s}$
and reference angle $\theta=\SI{1}{\degree}$.
For this source, the projected longitudinal FWHM sets
the Gaussian reference width,
$\Gamma_G=\Delta v_{\parallel}^{\mathrm{FWHM}}
\sin\theta/\lambda$.
This gives $(55.2,78.8,105.6)$ MHz for
\SI{791}{nm}, \SI{554}{nm}, and \SI{413}{nm}, respectively.
The Lorentzian FWHM values remain $(8.24,19.0,9.15)$ MHz
for the \SI{791}{nm}, \SI{554}{nm}, and \SI{413}{nm}
transitions, respectively.

For $^{133}\mathrm{Ba}$, we select $F'=1/2$ at
\SI{791}{nm}, $F'=3/2$ at \SI{554}{nm}, and
$F'=1/2$ at \SI{413}{nm}. With $^{132}\mathrm{Ba}$ outnumbering $^{133}\mathrm{Ba}$
by nearly 50:1, the selected \SI{791}{nm} component
gives a calculated excitation selectivity of approximately
99.85\% at $\theta=\SI{1}{\degree}$ and retains higher
selectivity over the angular range shown than the other
two transitions (Fig.~\ref{fig:selectivitycomparison}(e,f)).
In summary, the \SI{791}{nm} transition provides strong
first-step discrimination under both source conditions.
The resulting \SI{791}{nm} isotope and intermediate-state weights
supply the input to the relayed-selectivity calculation
in Sec.~\ref{sec:sup554}.

\section{First-principles calculations of isotope shifts and hyperfine structure}
\label{sec:first-principles-isotope-hfs}

\subsection{Electronic structure and hyperfine interaction}
We calculate the electronic structure of neutral Ba using the relativistic configuration-interaction
plus linearized coupled-cluster singles-and-doubles (CI+LCCSD)
method~\cite{Safronova2009CIAllOrder,Cheung2025PCI}. The calculation uses
the all-electron Dirac--Coulomb Hamiltonian and a $V^{N-2}$ Dirac--Fock
reference for the Xe-like closed core, where $N=56$ is the total
electron number of neutral Ba. Two valence electrons are treated
explicitly. For isotope mass number $A$, the nuclear charge
is distributed uniformly within a sphere of radius
$R_A=\sqrt{5/3}\,r_{\mathrm{rms},A}$, where
$r_{\mathrm{rms},A}$ is the root-mean-square nuclear charge radius.
The electronic state $\Phi_{\gamma J}$ is obtained by diagonalizing
\begin{equation}
 H_{\mathrm{eff}}=
 \sum_{i=1}^{2}\bigl[h_{\mathrm{DF}}(i)+\Sigma_1(i)\bigr]
 +\frac{1}{r_{12}}+\Sigma_2(1,2).
 \label{eq:fp-heff}
\end{equation}
Atomic units are used for the electronic-structure operators.
Here $h_{\mathrm{DF}}$ is the one-electron Dirac--Fock operator,
$r_{12}$ is the distance between the two valence electrons,
$J$ is the total electronic angular momentum, and $\gamma$
denotes the remaining electronic-state labels.
CI treats valence--valence mixing, while $\Sigma_1$ and $\Sigma_2$
describe core--valence correlations and core screening of the valence
interaction. We use pCI~\cite{Cheung2025PCI} with the hyperfine-operator
and recoil extensions specified below.

The radial basis comprises 50 B-splines of order 11 in a $60a_0$ cavity,
with radial orbital labels retained through $n=40$. The positive-energy
virtual space extends through $l=5$ and $500E_h$ for CI,
and through $l=7$ and $20000E_h$ for the coupled-cluster sums. Here $a_0$ and $E_h$ are the Bohr radius and Hartree energy,
$n$ labels radial orbitals, and $l$ is the orbital angular momentum. The all-order external space for $\Sigma_{1,2}$ contains the 24 orbitals
$6$--$9s$, $6$--$9p_{1/2,3/2}$, $5$--$8d_{3/2,5/2}$, and
$4$--$5f_{5/2,7/2}$. The remaining matrix elements are evaluated in
second-order many-body perturbation theory.

Hyperfine constants are evaluated at zero recoil. Following the
CI--coupled-cluster effective-operator framework~\cite{Tang2026Determination}, we
retain the full connected SD numerator and its common two-valence
normalization. We write the SD wave operator as
$\Omega_{\mathrm{SD}}=1+T_1+T_2+S_{cv}$, where $T_1$ and $T_2$
describe single and double core excitations, respectively, and
$S_{cv}$ excites one core and one valence electron.
The rank-$k$ ($k=1,2$) tensor expectation value is
\begin{equation}
 \langle T^{(k)}_0\rangle_{\gamma J}=
 \frac{\langle\Phi_{\gamma JJ}|[\Omega_{\mathrm{SD}}^{\dagger}
 T^{(k)}_0\Omega_{\mathrm{SD}}]_{\mathrm{conn}}|\Phi_{\gamma JJ}\rangle}
 {\langle\Phi_{\gamma JJ}|[\Omega_{\mathrm{SD}}^{\dagger}
 \Omega_{\mathrm{SD}}]_{\mathrm{conn}}|\Phi_{\gamma JJ}\rangle}.
 \label{eq:fp-connected-sd}
\end{equation}
Here $T^{(k)}_q$ is the electronic hyperfine tensor, with $k=1$
and $2$ corresponding to the magnetic-dipole and electric-quadrupole
interactions. $q$ labels the spherical component.
The state $\Phi_{\gamma JJ}$ has electronic angular-momentum
projection $M_J=J$, and ``conn'' denotes connected contractions. The numerator includes the linear and quadratic SD contractions, and
the denominator retains the identity and all connected norm terms,
including mixed core--valence contributions. Disconnected core factors
cancel in the ratio. The two-valence terms use the iterated $S_{cv}$
amplitudes, replacing the corresponding second-order two-particle
contribution. The ratio is evaluated without re-expanding the norm.
This construction includes core polarization, structural radiation,
and the associated normalization contributions within the retained SD
space. Nuclear magnetic and quadrupole distributions use the same
radius $R_A$. The point-multipole radial kernels are multiplied
by $\min[1,(r/R_A)^{2k+1}]$, where $r$ is the distance from the
nuclear center~\cite{Tang2026Determination}. The adopted nuclear parameters are listed in
Table~\ref{tab:fp-nuclear-inputs}.

\begin{table}[!t]
\centering
{\suppTableStyle
\begin{tabular}{cccrr}
\toprule
Isotope & $r_{\mathrm{rms},A}$ (fm) & $I_A$
& $\mu_A/\mu_N$ & $Q_A$ (b)\\
\midrule
$^{132}$Ba & 4.8303 & $0$   & $0$       & $0$\\
$^{133}$Ba & 4.8286 & $1/2$ & $-0.7719$ & $0$\\
$^{134}$Ba & 4.8322 & $0$   & $0$       & $0$\\
$^{135}$Ba & 4.8294 & $3/2$ & $0.8381$  & $0.153$\\
$^{136}$Ba & 4.8334 & $0$   & $0$       & $0$\\
$^{137}$Ba & 4.8314 & $3/2$ & $0.9375$  & $0.236$\\
$^{138}$Ba & 4.8378 & $0$   & $0$       & $0$\\
\bottomrule
\end{tabular}
}
\caption{Nuclear parameters used in the calculations.
Here $I_A$, $\mu_A$, and $Q_A$ are the nuclear spin,
magnetic dipole moment, and electric quadrupole moment of isotope $A$.
$\mu_N$ is the nuclear magneton and b denotes the barn.
Charge radii are taken from Ref.~\cite{Angeli2013Table},
magnetic moments from
Refs.~\cite{Stone2019Moments,Hoehle1976},
and quadrupole moments from
Ref.~\cite{Sahoo2013reliability}.}
\label{tab:fp-nuclear-inputs}
\end{table}

For each isotope, we obtain the magnetic-dipole and
electric-quadrupole hyperfine constants $A_\gamma$ and $B_\gamma$
by combining the electronic matrix elements with the nuclear
moments in Table~\ref{tab:fp-nuclear-inputs}, using the conventions
of Ref.~\cite{Tang2026Determination}.
Expressed in frequency units, these constants give the
first-order hyperfine shifts about the electronic centroid:
\begin{equation}
 \begin{aligned}
 \Delta\nu_{\gamma F}^{A}
 &=\frac{A_\gamma C_F}{2}\\
 &\quad+B_\gamma
 \frac{\tfrac34 C_F(C_F+1)-I_A(I_A+1)J(J+1)}
 {2I_A(2I_A-1)J(2J-1)}.
 \end{aligned}
 \label{eq:fp-casimir}
\end{equation}
where $F$ is the total atomic angular momentum and
$C_F=F(F+1)-I_A(I_A+1)-J(J+1)$.
The quadrupole term is omitted for $I_A<1$ or $J<1$.
For $^{135,137}$Ba and $J=1$, the interval used here is
$d_1=|\nu_{F=3/2}-\nu_{F=1/2}|=|3A_\gamma/2-9B_\gamma/4|$.
For $^{133}$Ba, the quadrupole term vanishes and the $F=1/2,3/2$
interval is $3|A_\gamma|/2$. The even isotopes have $I_A=0$
and no hyperfine splitting.

\subsection{Nuclear recoil and isotope shifts}
We obtain the mass-shift coefficients from a finite-field extension
of the same pCI workflow. In atomic units, the perturbed Hamiltonian
is $H(\lambda)=H_{\mathrm{DC}}+\lambda\mathcal R$, where
$H_{\mathrm{DC}}$ is the zero-recoil Dirac--Coulomb Hamiltonian
for the specified nuclear charge distribution.
The dimensionless auxiliary parameter $\lambda$ has the physical
value $m_e/M_A$, where $m_e$ and $M_A$ are the electron and
nuclear masses, respectively.
The low-order relativistic recoil operator is~\cite{Aleksandrov2015Finite}
\begin{equation}
 \mathcal R=\frac12\sum_{i,j}
 \left[\mathbf p_i\!\cdot\!\mathbf p_j
 -\mathbf D_i\!\cdot\!\mathbf p_j
 -\mathbf p_i\!\cdot\!\mathbf D_j\right].
 \label{eq:fp-recoil-operator}
\end{equation}
Here $i,j$ run over all electrons and $\mathbf p_i$ is the
momentum operator of electron $i$.
The $i=j$ and $i\ne j$ terms are the normal and specific mass shifts
(NMS and SMS), respectively. The finite-nucleus operator $\mathbf D$
is the normalized charge-density average of
$Z[\boldsymbol\alpha+(\boldsymbol\alpha\!\cdot\!\hat{\mathbf s})
\hat{\mathbf s}]/(2cs)$, where $\mathbf s$ joins a nuclear charge
element to the electron, $s=|\mathbf s|$, and
$\hat{\mathbf s}=\mathbf s/s$.
Here $Z=56$, $\boldsymbol\alpha$ is the vector of Dirac matrices,
and $c$ is the speed of light in atomic units.
The charge-density average uses the same uniform sphere
as the Dirac--Fock potential.

We first solve the static linearized Dirac--Fock response to the
one-electron recoil source. If $f'$ is the resulting Fock-matrix
derivative, the orbitals and reference
energies at each field are generated as
\begin{equation}
 \begin{split}
 \phi(\lambda)&=\phi(0)e^{\lambda X},\\
 \epsilon_p(\lambda)&=\epsilon_p(0)+\lambda f'_{pp},\\
 X_{pq}&=\frac{f'_{pq}}{\epsilon_q-\epsilon_p}\quad(p\ne q),\qquad X_{pp}=0.
 \end{split}
 \label{eq:fp-orbital-response}
\end{equation}
Here $\phi$ collects the one-electron orbitals as a row vector,
$p,q$ label orbitals, $\epsilon_p=\epsilon_p(0)$ is an unperturbed
orbital energy, and $X$ is the anti-Hermitian generator of the rotation.
The rotation acts within each relativistic angular
channel. Coulomb integrals, core and valence SD amplitudes,
$\Sigma_{1,2}$, and CI eigenstates are then recalculated in this
responding basis.

Let $\mathcal E_u^{A}(\lambda)
=E_u^{A}(\lambda)-E_g^{A}(\lambda)$ denote the
excitation energy from $g=6s^2\,{}^1S_0$ to $u=6s6d\,{}^3D_1$.
The energies $E_g^{A}$ and $E_u^{A}$ are evaluated without
hyperfine splitting. We evaluate the mass-shift coefficient
using a central finite difference at $\lambda=\pm\delta$,
where $\delta$ is the dimensionless field step:
\begin{equation}
 K_u^{A}\simeq\frac{m_e}{h}
 \frac{\mathcal E_u^{A}(+\delta)
       -\mathcal E_u^{A}(-\delta)}{2\delta}.
 \label{eq:fp-k-definition}
\end{equation}
Here $h$ is Planck's constant, and $K_u^{A}$ includes the recoil
contributions of both the ground and upper states.

\begin{table*}[!t]
\centering
{\suppTableStyle
\setlength{\tabcolsep}{6pt}
\begin{tabular}{ccccrrr}
\toprule
& & & & \multicolumn{3}{c}{Frequency offset (MHz)}\\
\cmidrule(lr){5-7}
Isotope & $F'$ & $F''$ & $J'\to J''$
& $\delta\nu_{u,A,F''}$ & $\delta\nu^{(2)}_{A,F'F''}$
& $\Delta_{554}^{\mathrm{th}}$\\
\midrule
$^{138}\mathrm{Ba}$ & -- & -- & $1\to1$ & 0 & 0 & $-5051.519$\\
$^{136}\mathrm{Ba}$ & -- & -- & $1\to1$ & 71.274 & $-37.226$ & $-5088.744$\\
$^{134}\mathrm{Ba}$ & -- & -- & $1\to1$ & 53.139 & $-69.161$ & $-5120.680$\\
$^{137}\mathrm{Ba}$ & $1/2$ & $1/2$ & $1\to1$ & 2308.519 & 5051.519 & 0\\
$^{137}\mathrm{Ba}$ & $1/2$ & $3/2$ & $1\to1$ & 992.527 & 3735.527 & $-1315.992$\\
$^{137}\mathrm{Ba}$ & -- & $5/2$ & -- & $-1179.785$ & -- & --\\
$^{135}\mathrm{Ba}$ & $1/2$ & $1/2$ & $1\to1$ & 2073.380 & 4458.980 & $-592.538$\\
$^{135}\mathrm{Ba}$ & $1/2$ & $3/2$ & $1\to1$ & 899.225 & 3284.825 & $-1766.693$\\
$^{135}\mathrm{Ba}$ & -- & $5/2$ & -- & $-1044.080$ & -- & --\\
\bottomrule
\end{tabular}
}
\caption{Calculated upper-state offsets and derived second-step
frequencies for the natural-Ba comparison.
The final column is referenced to the calculated
$^{137}\mathrm{Ba}$ $F'=1/2\rightarrow F''=1/2$ line.
For the odd isotopes, second-step offsets are listed for
excitation from $F'=1/2$.
The $F''=5/2$ rows list upper-state offsets only.
All additional allowed paths are constructed from these
upper-state levels and the intermediate-state positions in
Table~\ref{tab:sup137offsets}.
Values are rounded to three decimal places.
Calculations use unrounded offsets.}
\label{tab:sup554theory}
\end{table*}

The field shift is obtained directly from the zero-recoil excitation
energies, $\delta\nu_{u,\mathrm{FS}}^{A,138}
=[\mathcal E_u^{A}(0)-\mathcal E_u^{138}(0)]/h$.
The ground-to-upper hyperfine-component offset relative to $^{138}$Ba
is then
\begin{equation}
 \delta\nu_{u,A,F''}=
 \delta\nu_{u,\mathrm{FS}}^{A,138}
 +\frac{K_u^{A}}{M_A}
 -\frac{K_u^{138}}{M_{138}}
 +\Delta\nu_{uF''}^{A},
 \label{eq:fp-upper-offset}
\end{equation}
where $F''$ labels the upper-state hyperfine component and
$\Delta\nu_{uF''}^{A}$ is its shift from the electronic centroid,
evaluated using the zero-recoil full-connected-SD constants
$A_\gamma,B_\gamma$ for the upper state.
The nuclear masses $M_A$ are derived from the AME2020 atomic
masses~\cite{Wang2021AME}.
These upper-state offsets are combined
with the measured or predicted $6s6p\,{}^3P_1^o$ offsets to construct
the second-step frequencies in Sec.~\ref{sec:sup554}.

\section{Calculated second-step spectrum and relayed selectivity}
\label{sec:sup554}

We combine the calculated $6s6d\,{}^3D_1$ structure with
the \SI{791}{nm} component frequencies to obtain the
second-step resonance frequencies and evaluate the
additional isotope discrimination. First, we obtain the second-step frequency offsets relative
to $^{138}\mathrm{Ba}$ by subtracting the intermediate-state
offsets from the upper-state offsets:
\begin{equation}
\delta\nu^{(2)}_{A,F'F''}
=\delta\nu_{u,A,F''}-\delta\nu_{l,A,F'}.
\label{eq:supSecondShift}
\end{equation}
Here $\delta\nu_{u,A,F''}$ and $\delta\nu_{l,A,F'}$
are the ground-to-upper and ground-to-intermediate frequency
offsets relative to the corresponding $^{138}\mathrm{Ba}$
transitions.
The upper-state values include the field shift, nuclear recoil,
and hyperfine structure calculated in
Sec.~\ref{sec:first-principles-isotope-hfs}.
The intermediate-state values are measured for the natural-Ba
comparison (Table~\ref{tab:sup137offsets}) and estimated by
King plots and hyperfine scaling for $^{132,133}\mathrm{Ba}$
(Table~\ref{tab:sup133offsets}).
Table~\ref{tab:sup554theory} lists the natural-Ba results,
with one common frequency translation aligning the calculated
$^{137}\mathrm{Ba}$ $F'=1/2\rightarrow F''=1/2$ line
with its fitted center in Fig.~3 of the main text.

Second, we extract the four measured resonance centers
with a joint zero-baseline Voigt fit.
The profiles share one Gaussian FWHM, with independent
Lorentzian widths, centers, and amplitudes.
The fit uses inverse-variance weights from the standard
errors of five-trial means and gives a shared second-step
Gaussian FWHM of \SI{56.9}{MHz}.
Quoted uncertainties are statistical errors from the unscaled
fit covariance, including correlations between centers
for frequency differences.

The calculated $F''=1/2$--$3/2$ separations are
\SI{1316.0}{MHz} for $^{137}\mathrm{Ba}$ and
\SI{1174.2}{MHz} for $^{135}\mathrm{Ba}$, compared with
fitted values of $1305(2)$ and $1157(3)~\mathrm{MHz}$.
The calculated separations between the two isotopes are
\SI{592.5}{MHz} and \SI{450.7}{MHz} for the
$F'=1/2\rightarrow F''=1/2$ and
$F'=1/2\rightarrow F''=3/2$ branches, respectively,
compared with $595(2)$ and $447(3)~\mathrm{MHz}$ from the fits.
These comparisons support the line assignments and isotope
separations underlying the second spectral filter.

Finally, we carry the first-step spectral weights $p_{A,F'}=a_As_{A,F'}\mathcal V_1
(\nu_1-\nu^{(1)}_{A,F'})$ into the sequential response in Eq.~(3) of the main text.
Here $\nu_1$ and $\nu^{(1)}_{A,F'}$ are the \SI{791}{nm}
laser and component frequencies, and $\mathcal V_1$ is
the \SI{791}{nm} peak-normalized Voigt
profile.
Each weight is the abundance-weighted contribution of one
intermediate-state component, so
$R_A^{(1)}=\sum_{F'}p_{A,F'}$. To compare the spectral filtering of the two steps,
we average their excitation strengths over equally
populated initial magnetic sublevels.
For $J'=1\rightarrow J''=1$ excitation,
the normalized second-step strengths are
$t_{A,F'F''}
=3(2F''+1)
\left\{
\begin{smallmatrix}
1&F''&I_A\\
F'&1&1
\end{smallmatrix}
\right\}^{2}$, where $I_A$ is the nuclear spin and the braces denote a
Wigner $6j$ symbol.
These strengths satisfy $\sum_{F''}t_{A,F'F''}=1$.
For excitation from $F'=1/2$ to $F''=1/2$ and $3/2$,
they are $(1/6,5/6)$ in $^{135,137}\mathrm{Ba}$
and $(2/3,1/3)$ in $^{133}\mathrm{Ba}$.
All allowed paths from every intermediate-state component
are included in the sequential response.

\begin{figure*}[!t]
\centering
\begin{minipage}[t]{\columnwidth}
    \centering
    \makebox[\linewidth][l]{\textbf{(a)}}
    \includegraphics[width=\linewidth]
    {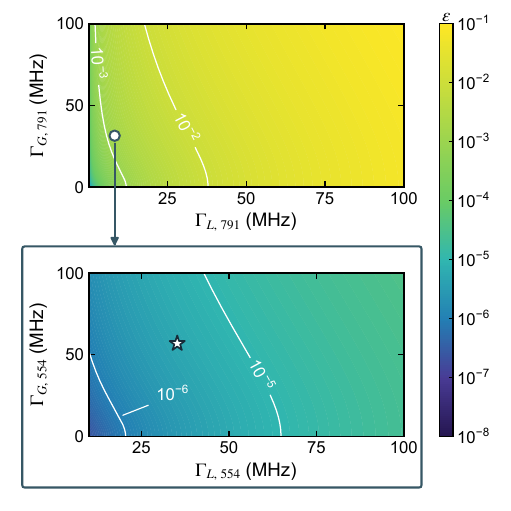}
\end{minipage}
\hfill
\begin{minipage}[t]{\columnwidth}
    \centering
    \makebox[\linewidth][l]{\textbf{(b)}}
    \includegraphics[width=\linewidth]
    {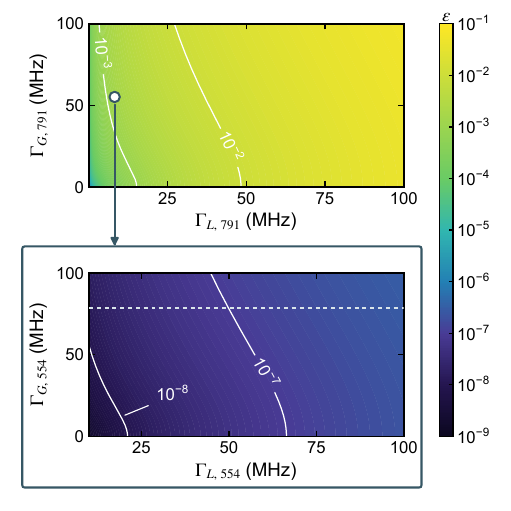}
\end{minipage}
\caption{Calculated relayed selectivity for
(a) $^{135}\mathrm{Ba}$ from natural-abundance barium and
(b) the synthetic $^{133}\mathrm{Ba}$ source.
The upper and lower maps show $\epsilon_1^{(A_t)}$ and
$\epsilon_2^{(A_t)}$ on logarithmic color scales.
The lasers address the target $F'=1/2$ first-step component
and $F'=1/2\rightarrow F''=1/2$ second-step component.
The arrow and frame indicate that the lower map retains
the first-step weights at the circle while varying the
second-step widths.
In (a), the circle and star use the $^{137}\mathrm{Ba}$
reference linewidths given in the text.
In (b), the source contains 2\% $^{133}\mathrm{Ba}$,
97\% $^{132}\mathrm{Ba}$, and 1\% $^{134}\mathrm{Ba}$.
The circle uses the ablation-source parameters, and
the horizontal dashed line marks
$\Gamma_{G,554}=\SI{78.8}{MHz}$.}
\label{fig:supRelay135133}
\end{figure*}

For target isotope $A_t$, we define the unwanted-isotope
excitation fraction after one or two steps as
\begin{equation}
\epsilon_k^{(A_t)}
=\frac{\sum_{A\ne A_t}R_A^{(k)}}{\sum_A R_A^{(k)}},
\qquad k=1,2.
\label{eq:supRelayOtherTargets}
\end{equation}
For the \SI{791}{nm} response, this gives
$\epsilon_1^{(A_t)}=1-\eta_{A_t}$,
linking the calculation to Sec.~\ref{sec:supcomparison}.
For $A_t=137$, $\epsilon_k^{(137)}$ recovers the
main-text $\epsilon_k$.
For each target, the \SI{791}{nm} laser addresses the
$F'=1/2$ component selected in Sec.~\ref{sec:supcomparison}.
The \SI{554}{nm} laser addresses the calculated
$F'=1/2\rightarrow F''=1/2$ resonance.
Under the reference conditions specified below, this branch
gives a smaller $\epsilon_2^{(A_t)}$ than the $F''=3/2$ branch.

For $^{137}\mathrm{Ba}$ in natural-abundance source, we use the natural
abundances of Sec.~\ref{sec:supcomparison} and the reference
first-step widths
$(\Gamma_{L,791},\Gamma_{G,791})=(8.24,31.4)$ MHz,
giving $\epsilon_1^{(137)}=5.94\times10^{-4}$.
The fitted $^{137}\mathrm{Ba}$
$F'=1/2\rightarrow F''=1/2$ profile sets the second-step
widths to $(\Gamma_{L,554},\Gamma_{G,554})=(35.2,56.9)$ MHz.
This gives $\epsilon_2^{(137)}\simeq6.8\times10^{-7}$,
an approximately 870-fold reduction (Fig.~4 of the main text).
Retuning both steps to $^{135}\mathrm{Ba}$ at the same
abundances and linewidths gives
$\epsilon_1^{(135)}=1.56\times10^{-3}$ and
$\epsilon_2^{(135)}\simeq5.3\times10^{-6}$,
an approximately 290-fold reduction
(Fig.~\ref{fig:supRelay135133}(a)).

For the synthetic $^{133}\mathrm{Ba}$ source, we retain
the conditions specified in Sec.~\ref{sec:supcomparison}.
The second-step frequencies follow
Eq.~\eqref{eq:supSecondShift} and are listed in
Table~\ref{tab:sup133SecondStep}.
The selected $^{133}\mathrm{Ba}$ line lies below the
$^{132}\mathrm{Ba}$ and $^{134}\mathrm{Ba}$ lines by
\SI{5020.2}{MHz} and \SI{5044.7}{MHz}, respectively,
providing a second spectral filter against the dominant
source isotopes. For this source, the reference first-step widths
$(\Gamma_{L,791},\Gamma_{G,791})=(8.24,55.2)$ MHz
give $\epsilon_1^{(133)}=1.52\times10^{-3}$.
For the second step, we keep the Gaussian width fixed
and vary $(\Gamma_{L,554},\Gamma_{G,554})$
from $(10,78.8)$ to $(100,78.8)$ MHz.
The resulting $\epsilon_2^{(133)}$ increases from
$1.36\times10^{-8}$ to $2.93\times10^{-7}$,
corresponding to more than $5\times10^3$ additional
suppression throughout this range
(Fig.~\ref{fig:supRelay135133}(b)).
In summary, the separated second-step resonances extend
relayed discrimination to both natural-abundance barium
and a synthetic source dominated by neighboring isotopes.

\begin{table}
\centering
{\suppTableStyle
\begin{tabular}{cccr}
\toprule
Isotope & $F'$ & $F''$ & $\delta\nu^{(2)}_{A,F'F''}$ (MHz)\\
\midrule
$^{132}\mathrm{Ba}$ & -- & -- & $-93.655$\\
$^{133}\mathrm{Ba}$ & $1/2$ & $1/2$ & $-5113.822$\\
$^{133}\mathrm{Ba}$ & $1/2$ & $3/2$ & $-1886.510$\\
$^{133}\mathrm{Ba}$ & $3/2$ & $1/2$ & $-853.442$\\
$^{133}\mathrm{Ba}$ & $3/2$ & $3/2$ & $2373.869$\\
$^{134}\mathrm{Ba}$ & -- & -- & $-69.161$\\
\bottomrule
\end{tabular}
}
\caption{Second-step offsets for the synthetic
$^{133}\mathrm{Ba}$ source, relative to the
$^{138}\mathrm{Ba}$ second-step transition.
All allowed paths are listed.
The upper-state offsets come from
Sec.~\ref{sec:first-principles-isotope-hfs} and the
intermediate-state offsets from Table~\ref{tab:sup133offsets}.
For even isotopes, the transition is $J'=1\rightarrow J''=1$.
Values are rounded to three decimal places.
Calculations use unrounded offsets.}
\label{tab:sup133SecondStep}
\end{table}

\section{Intensity dependence of the final ionization step}
\label{sec:sup532}

We also characterize the final ionization step by varying the
\SI{532}{nm} peak intensity while the \SI{791}{nm} laser is held
near its loading maximum and the \SI{554}{nm} laser is tuned
to the $F'=1/2\rightarrow F''=1/2$ loading resonance. As shown in
Fig.~\ref{fig:532power}, the loading rate increases approximately
linearly over the measured range from
\SI{2.2}{kW.cm^{-2}} to \SI{13.4}{kW.cm^{-2}}. A weighted linear fit gives a slope of
$0.37(3)\,(\mathrm{ions/s})/(\mathrm{kW/cm^2})$. No saturation is observed
over this range, consistent with operation in the unsaturated regime
of the final one-photon ionization step from $6s6d\,{}^3D_1$.
The use of direct photoionization at \SI{532}{nm} is not necessarily
the optimal implementation of the final step. Starting from the
same $6s6d\,{}^3D_1$ intermediate state, a suitably chosen
odd-parity autoionizing resonance could in principle provide a larger
photoionization cross section. Literature estimates place the
non-resonant photoionization cross section of $6s6d\,{}^3D_1$ on the order of
$10^3~\mathrm{Mb}$ \cite{Kallenbach1989}, while strong autoionizing
resonances in Ba can further improve the ionization rate but on the same order of magnitude
\cite{Wall2026}.
A detailed calculation, however, depends on the
specific autoionizing state accessible from $6s6d\,{}^3D_1$ and on
its linewidth relative to the Doppler width, which is beyond the scope of the work.

\begin{figure}[h]
\centering
\includegraphics[width=\columnwidth]{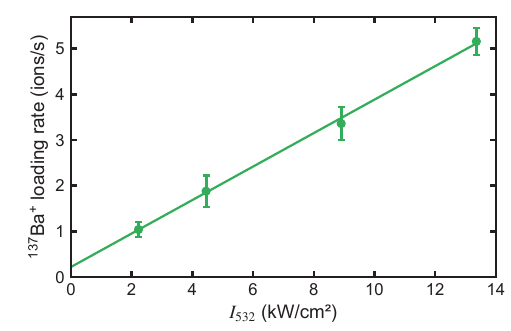}
\caption{Loading rate versus peak intensity of the \SI{532}{nm}
ionization beam, with the \SI{791}{nm} laser near its
$^{137}\mathrm{Ba}$ loading maximum and the \SI{554}{nm} laser on the
$F'=1/2\rightarrow F''=1/2$ resonance. Error bars denote one standard
error. The solid line is a weighted linear fit with a free
intercept and slope
$0.37(3)\,(\mathrm{ions/s})/(\mathrm{kW/cm^2})$.
The highest measured intensity corresponds to approximately
\SI{300}{mW}, comparable to the routine global Raman power.
No saturation is resolved over the measured range.}
\label{fig:532power}
\end{figure}
}
  {}

\end{document}